\documentclass{llncs}
\usepackage{graphicx} 
\usepackage{url}
\usepackage[utf8]{inputenc}
\usepackage[english]{babel}
\usepackage{amsmath}
\usepackage{amsfonts}
\usepackage{amssymb}
\usepackage{booktabs}

\begin{document}

\mainmatter              % start of the contributions
\title{Query Expansion for Survey Question Retrieval in the Social Sciences}
\author{Nadine Dulisch\inst{1} \and Andreas Oskar Kempf\inst{2} \and Philipp Schaer\inst{1}
\thanks{Authors are listed in alphabetical order.}}
\institute{
GESIS -- Leibniz Institute for the Social Sciences, 50669 Cologne, Germany\\
\email{firstname.lastname@gesis.org}\\
\and
ZBW -- German National Library for Economics, 20354 Hamburg, Germany\\
\email{a.kempf@zbw.eu}
}

\maketitle  

\begin{abstract}
In recent years, the importance of research data and the need to archive and to share it in the scientific community have increased enormously. This introduces a whole new set of challenges for digital libraries. In the social sciences typical research data sets consist of surveys and questionnaires. In this paper we focus on the use case of social science survey question reuse and on mechanisms to support users in the query formulation for data sets. We describe and evaluate thesaurus- and co-occurrence-based approaches for query expansion to improve retrieval quality in digital libraries and research data archives. The challenge here is to translate the information need and the underlying sociological phenomena into proper queries. As we can show retrieval quality can be improved by adding related terms to the queries. In a direct comparison automatically expanded queries using extracted co-occurring terms can provide better results than queries manually reformulated by a domain expert and better results than a keyword-based BM25 baseline.
\end{abstract}

 % Not required for Proceedings

\section{Introduction}
Digital libraries in academia increasingly include research data sets \cite{dallmeier-tiessen_integrating_2014}. In order to facilitate data reuse, a retrieval infrastructure for research data needs to be built up \cite{hyman_use_2006}. Taking the example of quantitative social science research, research data for the purpose of reuse mostly consist of survey data, i.e. data collected to capture attitudes and behaviors as well as factual information of a population or population groups. The core method of data collection in survey methodology is the questionnaire. It provides the basis on which respondents' answers are converted into data that can be analyzed statistically. With respect to the reuse scenario of survey data, it is important to keep in mind that other researchers are less interested in the entire survey. Rather they are looking for data on a single studied phenomenon or social construct and how it is translated into individual questions or items as part of the questionnaire. 

Regarding the development of a retrieval infrastructure for survey question reuse it is important to keep in mind that in the majority of cases the measured social construct can hardly be derived directly from the question text itself (which is explained in more detail in section \ref{sec:surveydata}). They need to be broken down into measurable properties, this way getting operationalized in the form of survey questions or items as part of a questionnaire -- the so-called measuring instrument. It is this operationalization process which is the main reason for the great interest in concrete survey questions and which at the same time constitutes a major challenge for the development of retrieval services for the reuse of survey questions. Not only does the exact wording of a question determine whether a survey question really is suitable to allow for conclusions on a phenomenon and therefore is a valid measuring instrument. It also determines whether results of one survey could be compared with results of another survey which pretends to investigate the same phenomenon. Researchers then could resort to questions or whole item batteries that have already been developed by other researchers and used in various studies. Tested and established measuring instruments for the social sciences can be found in special scientific databases like question banks. Provided that a full text documentation of survey data questionnaires exists researchers could find exactly those measuring instruments they need to operationalize their own research interest. 

Based on an analysis of the search log files of ZACAT\footnote{\url{http://zacat.gesis.org/webview/}}, the GESIS online study catalogue system, potential re-users of survey question usually search with keywords related to the phenomenon they are interested in. This stands in contrast to the actual information contained in ZACAT where usually the actual question text itself is stored whereas the underlying social construct could only be found in the study description. For this reason a mere string-based search query is not enough to find well-established survey questions suitable to measure phenomena directly linked to one’s own research interest. In this paper we want to find out whether a query expansion system is a suitable tool to support the retrieval of survey questions. This retrieval support is motivated by the idea to improve the reuse of survey questions across different studies and questionnaires.

\section{Social Science Survey Data}
\label{sec:surveydata}
\begin{figure}[t]
\centering
\includegraphics[width=\textwidth]{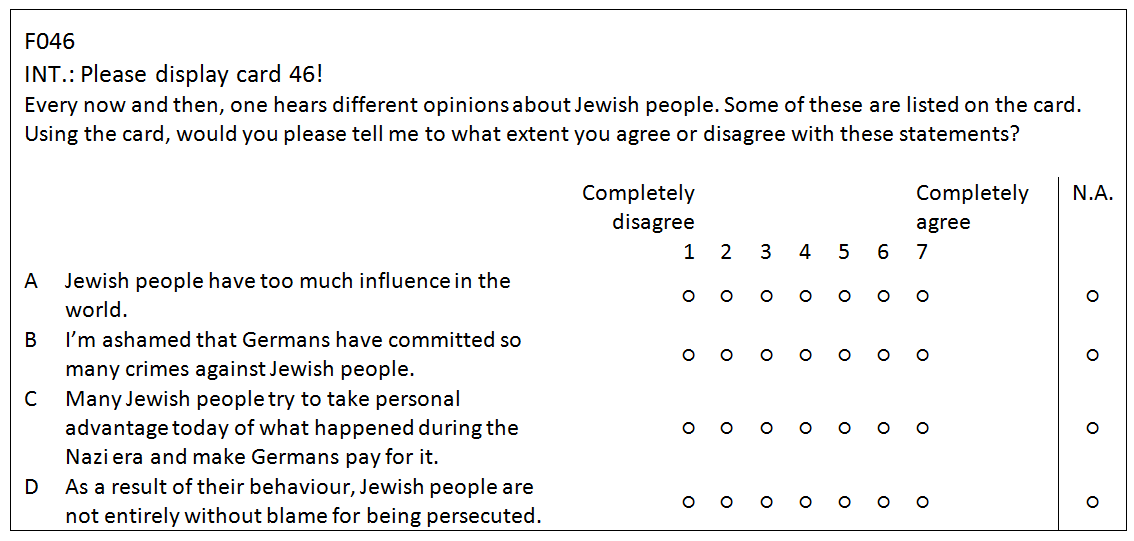}
\caption{Excerpt from the German General Social Survey (ALLBUS) 2012}
\label{fig:survey}
\end{figure}

Social science surveys essentially consist of numerical data. The data files are usually composed of tables which entail numbers or codes, which represent the values of respondents’ answers to a survey question. It is for this reason why subject content of a survey can hardly be concluded by the dataset itself, but rather from other pieces of information linked to the study the data had been collected for. It could only be derived from the study as a whole \cite{friedrich_making_2014}. In general, three different levels of survey data structure can be distinguished. On the study level information about the general content of a study is provided. It includes, for example, information about the research fields of the data producers as well as the codebook. The narrower level is the variable level which gives detailed information on the phenomena under study. It contains the different variables which reflect the various dimensions of the definition of the studied phenomenon laying the ground for the entire formulation of the questionnaire. The third level is the question level. It contains the concrete questions or items the respondents have to deal with, which could be formulated rather differently, ranging from question format and statements (see figure~\ref{fig:survey}) to tasks which need to be solved. In this article we focus on the questionnaire, which is at the core of each survey.

A constitutive part of questionnaire design is the operationalization process which stands for the translation of a research construct (e.g. antisemitism) into measurable units. The researcher identifies the different dimensions of a construct and defines it referring to relevant research literature, earlier studies or even his or her own qualitative pre-studies in the field. He or she then derives measurable aspects out of this definition which can be included in the different survey questions (e.g. agreement/disagreement with the statement: ``Jewish people have too much influence in the world''). So the underlying research construct is encoded, which is why it usually cannot be extracted or derived out of the concrete question text. Only less complex social constructs or manifest variables as for example demographic variables like sex and age or the level of education usually appear in a questionnaire in their literal form.

The reuse scenario we focus on is based on the operationalization process as part of every questionnaire design. The user group we have in mind are social scientists who are planning to create their own surveys and who want to know how to put their research interest into concrete survey questions or items as part of a questionnaire. To the best of our knowledge there is no literature on the concrete search and information behavior of social scientists during the questionnaire design process. Information services like ZIS\footnote{\url{http://www.gesis.org/en/services/data-collection/zisehes/}} illustrate the importance of documentation of social science measurement instruments. The problem at hand is a typical information seeking problem where subject information resources (survey questions) have to be retrieved from a database by a search engine. Similar to other retrieval problems, we have to deal with vagueness and ambiguity of human language (see the following section on related work). This problem gets even more pronounced given the fact that question texts, as well as search queries in survey question databases, tend to be rather short. Typical survey questions in our data set (see section \ref{sec:datasets}) are less than 100 characters long and typical search queries contain less than two words. 

\section{Related Work}
\label{sec:relatedwork}
A typical problem that arises during every search-based retrieval task (in contrast to browsing or filter-based tasks) is the so-called language or vocabulary problem \cite{furnas_vocabulary_1987}: During the formulization of an information need, a searcher can (in theory) use the unlimited possibilities of human language to express him- or herself \cite{blair_information_2003}. This is especially true when expressing information needs in the scientific domain using domain-specific expressions that are very unique and context-sensitive. Every scientific community and discipline has developed its own special vocabulary that is not commonly used by other researchers from other domains. With regard to survey question retrieval, this problem is even more pronounced as it is likely that the underlying topic of a survey question is not directly represented in the question text. In this special setting of short textual documents and a very domain-specific content, this long-known problem becomes even more pronounced.

Hong et al. apply four methods for microblog retrieval \cite{hong_query_2011}: query reformulation, automatic query expansion, affinity propagation as well as a combination of these techniques. To reformulate the query hashtags are extracted from tweets and used as additional information for the query. Furthermore, every two consecutive words of the query are grouped and added to the query. A relevance feedback model is used for automatic query expansion. %\cite{lavrenko_relevance_2001}
The respective top ten terms of the top ten documents are selected. The affinity propagation approach is implemented by using a cluster algorithm to group tweets. %\cite{frey_clustering_2007}
The idea behind this is that the probability of tweets being relevant is higher for those, which are similar to relevant tweets. It is shown that automatic query expansion is a very effective method, while affinity propagation is less successful. 

Microblogging services like Twitter also face the vocabulary problem for short texts. A tweet consists of up to 140 characters, while the question texts used in this work have an average length of 83.57 characters. The latest research in the field of microblog retrieval is therefore relevant for the problem at hand. For instance, pseudo-relevance feedback \cite{miyanishi_improving_2013} and document expansion \cite{efron_improving_2012} are common approaches to address the vocabulary problem \cite{carpineto_survey_2012}.
Jabeur et al. analyze two approaches for microblog retrieval \cite{jabeur_irit_2013}. The first approach uses a retrieval model based on Bayesian networks. The influence of a microblogger as well as the temporal distribution of search terms are included in the calculation of the relevance of a tweet. Here, only the usage of topic-specific features improved the results. In the second approach, query expansion (pseudo-relevance feedback) and document expansion methods are implemented. Tweets obtained by these approaches are merged. Additionally, those tweets are extended by contained URLs. Final scores are calculated by applying Rocchio-expansion as well as using the vector space model. Document expansion combined with vector space model improves retrieval results. Automatic query expansion does not increase recall, but significantly increases precision.

To surpass the language problem in digital libraries tools like thesauri and classifications try to control the vagueness of human language by defining a strict rule set and controlled vocabularies. These tools can help in the query formulation phase by actively supporting users in expressing their information need. A wide range of possible query expansion and search term recommender techniques are known in the information retrieval community \cite{schaer_applied_2013}. These techniques can be categorized (1) global techniques that rely on the analysis of a whole collection, and (2) local techniques that emphasize the analysis of the top-ranked documents \cite{xu_improving_2000}. While local techniques generally outperform global techniques, global techniques cannot be applied. In web search engines, the use of query suggestion systems is common, however, the situation in digital library systems is different \cite{luke_framework_2013}. This holds true especially in the very special setting of survey question retrieval. In digital library systems the use of knowledge organization systems is common practice. Typically, entire collections are indexed with controlled terms from a domain-specific thesaurus or a classification system. A query expansion system might try to suggest terms that are closely related to both the users initial query term, as well as the knowledge organization system. While theoretically any kind of metadata may be recommended, the most promising approaches \cite{gradmann_novel_2011} use terms from a knowledge organization system. 

Commercial tools like Colectica or QuestionPro are software packages that allow questionnaire designers to reuse questions. These tools include so-called question banks that store previously used questions. These question banks only allow basic string-based queries and are therefore not suitable for supporting users in the best possible way.
Indexing of research data with controlled vocabularies has begun only recently. An overhauled version of Nesstar, the software system for research data publishing and online analysis owned by the Norwegian Social Science Data Services (NSD), provided an indexing function on the variable level. Initially intended for indexing of research data of social science data archive members of the European Social Science Data Archives Consortium (CESSDA), it has not been implemented so far. Nevertheless, the CESSDA consortium is planning to establish a database for social science survey questions using the European Language Social Science Thesaurus (ELSST). Organizations like the DDI Alliance highly advocate the reusability and the exchange of survey metadata and proposes to use the DDI metadata standard. 
%ODESI is an example for such a DDI-driven research data platform that contains over 3,539 datasets from the social sciences, with an additional 13,627 survey records.

\section{Test Collection Construction and Experimental Setup}
\label{sec:datasets}
In order to conduct a careful and considerable evaluation on the problem of survey question retrieval, we chose to implement a TREC-style evaluation setup with (1) a document corpus, (2) a set of topics, and (3) a set of relevance assessments corresponding to the set of topics.

The document corpus was extracted from the ALLBUS (German General Social Survey) and SOEP (German Socio-Economic Panel) questionnaires. It contained 16,764 question- and sub-question texts (see figure~\ref{fig:survey} for an example). The textual information was short, compared to normal TREC-style documents like newspaper articles or web pages. The question texts contained in the corpora had an average length of only 83.57 characters. We chose the ALLBUS and SOEP data sets as both are relevant and domain-specific questionnaires that are used in thousands of social science publications worldwide. 

We extracted the topic set from log files (between 2012-01-01 and 2013-06-02) of the social science data catalogue ZACAT. The log files of ZACAT were chosen because these queries represent real-world usage patterns of scientists who are looking for questionnaires and survey data sets. Typical TREC topic sets consist of at least 50 topics so we extracted 60 queries from the log files. Since the corpus of question texts consists of German texts solely and the ZACAT queries were predominantly formulated in English, the queries were translated into German (official translation from ALLBUS/SOEP were used if applicable).
%ZACAT allows to search within questionnaires as well as within the question texts they contain. It is used by scientists who search for questionnaires and survey data sets. 
The frequencies of the log entries show a typical power-law-like skewed distribution. Therefore to select the set of topics the log entries were arranged in descending order by their frequency. Queries were grouped according to their frequency (high = more than 10 log entries; medium = 10 log entries which was the mean value over all log entries; low = 1 log entry).
To gather a good mix of different kinds of query topics we selected random log entries from the different groups: 27 common entries (high frequencies), 17 out of the medium frequent and 16 from special (low frequent) entries. Most of the log entries (mostly the high frequency ones) were short keyword queries (1.97 terms on average).
%The final topic set consists of 60 topics, 27 with high (e.g. ``trust'', ``democracy'', ``religion''), 17 with medium (e.g. ``personality'', ``justice'', ``ideology'') and 16 with low frequency (e.g. ``work life'', ``life satisfaction'', ``job quality''). 

%The relevance assessments for each topic-question text pair were conducted with the aid of an assessment tool. The assessment tool provides a graphical user interface that displays all not yet assessed question texts for each topic as well as all already assessed question text to allow for reassessment.

The ground truth was composed out of a set of 12,190 relevance assessments based on a three level rating system (not relevant, (partially) relevant, and very relevant). The relevance assessments were conducted manually by one single person familiar with the domain and with a set of rules and guidelines for the assessments. An example for such a guideline for the topic ``democracy'' was: 
\begin{itemize}
\item very relevant: direct questions concerning democracy (e.g. ``How content are you with the present state of democracy in Germany?'');
\item relevant: political questions concerning democracy in a broader sense (e.g. party system, election, etc.);
\item not relevant: everything else. 
\end{itemize}
 %The assessment of a topic-question text pair can be done by clicking on the corresponding radio button. The ground truth can be exported into the qrel-format, which is the format used by our evaluation software (TREC\_EVAL). We As input, the TREC\_EVAL software needs the relevance assessments (in form of the qrel-file) as well as the retrieval results in form of a text file. This text file contains information like topic-id, document-id (of the question text), rank and score. Values for various effectiveness measures are calculated automatically using this information.
%
%\subsection{Evaluation Measures}
\noindent
All retrieval experiments were conducted using Lucene with BM25 ranking and language dependent settings (stemmers, stop words, etc.). In all experiments we report on recall at n (R@5 and R@10) and on nDCG at n (nDCG@5 and nDCG@10) calculated using TREC\_EVAL. In our setup we understand recall as fraction of the research questions that are relevant to the query and are successfully retrieved within the first n results in the result list. Normalized discounted cumulative gain (nDCG) is a measure of retrieval quality for ranked lists that, in contrast to precision, makes use of graded relevance assessments \cite{jarvelin_cumulated_2002}. NDCG is computed as follows:
\begin{equation}
\text{nDCG} = Z_i \sum\limits_{j=1}^{R} \frac{2^{r(j)}-1}{\log(1+j)}
\end{equation}
\noindent
$Z_i$ is a constant to normalize the result to the value of 1. $r(j)$ is an integer representing the relevance level of the result returned at rank j where R is the last possible ranking position. For our experiments the relevance levels are 0 = irrelevant; 1 = relevant and 2 = very relevant. NDCG@n is a variation of this calculation where only the top-n results are considered. We use nDCG@5 and nDCG@10. The same two levels are chosen for the calculation of recall at n. Here we simply count the relevant documents (relDocs) among the first 5 or 10 results and divide them by the actual level:
\begin{equation}
R@n = \frac{|\text{relDocs}|}{n}
\end{equation}
\noindent
Since all our retrieval results are ranked we focus on the top five and ten results for our evaluation. This is done to simulate the needs of a real world user who is used to inspect only the first few results in a ranked list. All results are tested for statistical significance using a paired t-test. 

\section{Query Expansion for Survey Question Retrieval}
\label{sec:methods}
For each approach, queries are expanded as follows: A query consists of one or more clauses which in turn consist of terms (or phrases) and Boolean operators (for query syntax also see the Apache Lucene documentation). For each query term, the corresponding expansion terms are retrieved and connected with the query term by OR-operation. An example query for two query terms $qt_1$ and $qt_2$ with their corresponding alternative query terms $qt_{1,alt}$ would look like:
\begin{equation*}
(qt_1 \: \text{OR} \: qt_{1,alt} \: \text{OR} \dots)\: \text{AND/OR} \: (qt_2 \: \text{OR} \: qt_{2,alt} \: \text{OR} \dots)
\end{equation*}
\noindent
For evaluation, we expanded the unprocessed queries from the query logs in order to ensure a realistic setting. 

\subsection{Thesaurus-Based Expansion}
Thesauri are tools to surpass the vocabulary problem: natural language allows expressing things in various ways. One word can have the same (synonyms) or different meaning (polysemes). Thesauri contain various associations and relationships between terms in the form of synonyms, associations, etc. By extending the query with these related terms, documents can be found which do not contain the terms of the original query, but are still relevant with respect to the information need of the user. Since thesauri are not developed automatically but by creating relations manually between terms, this is an intellectual approach. The first approach involves two different thesauri: (1) the \emph{Open Thesaurus} (OT), a general natural language and community-based thesaurus, (2) and the \emph{Thesaurus for the Social Sciences} (TSS), a domain-specific thesaurus developed by a small editorial group of domain experts \cite{zapilko_thesoz:_2013}.

The Open Thesaurus contains German synonyms, hypernyms, hyponyms and associations. It does not consist of a domain-specific language and is licensed under the LGPL (GNU Lesser General Public License) License. The TSS contains descriptor-based synonyms, broader-, narrower and related terms as well as preferred terms.
For expanding the queries with the Open Thesaurus we limit the terms to synonyms as well as associations. With TSS we limit the terms to synonyms and related/preferred terms.
The aim is to reduce the number of suggested expansion terms. As there is no criterion to determine the ``degree of relatedness'' to the query term, any limited selection would have been too random. Therefore, all determined expansion terms are added to the original query (by logically linking them to the original term with an OR function). These are 19.49 terms on average, which are retrieved from a database. Furthermore, to ensure comparability between both thesauri approaches, similar kinds of relationships have to be used (synonyms, associations/related terms). The goal of using a general thesaurus as well as a domain-specific thesaurus is to compare natural language expansion to domain language expansion.

\subsection{Co-occurrence-Based Expansion}
As described in the previous section, the thesaurus-based query expansion is an intellectual approach. Statistical approaches to determine terms for expansion have proven to be more applicable \cite{carpineto_survey_2012}. In this paper the statistical method of co-occurrence analysis has been tested. Co-occurrence analysis is a well-established approach, which is for instance used in natural language processing or to support manual coding of qualitative interviews \cite{brent_feeling_2003}. It serves the purpose of the analysis of term-term relationships. It is assumed that terms, which often occur within the same context, are associated with each other and are, for example, similar in meaning.

Since each similarity measure performs differently depending on the data we used two different metrics: logarithmic Jaccard index, also known as the Jaccard similarity coefficient and cosine similarity.
We define the logarithmic Jaccard index as follows: 
\begin{equation}
J_{log}(x,y) = \frac{\log(df_{xy})}{\log(df_x + df_y - df_{xy})}
\end{equation}
\noindent
The cosine similarity is calculated as follows: 
\begin{equation}
c(x,y)= \frac{df_{xy}}{\sqrt{df_x + df_y}}
\end{equation}
\noindent
$df_x$ and $df_y$  are the document frequencies of the terms x and y, thus the number of documents these terms occur in. $df_{xy}$ is the number of documents which contain x as well as y. Jaccard index corresponds to intersection/union.

As we want our co-occurrence-based method to be comparable to our previous approach that was based on a thesaurus we again focus on domain-specific vocabularies. For this purpose, we use the social science literature database SOLIS with more than 450,000 literature references from the social sciences. Each reference includes title, abstract and controlled keywords from the TSS. Our system calculates the semantic relatedness between any free text such as titles or abstracts and controlled terms (TSS-terms) for an entire document corpus (stop word removal and stemming is applied). Using co-occurrence measures like Jaccard index or cosine similarity we calculated term suggestions taken from the TSS for every query term. As an example, users who are looking for the string ``youth unemployment'' in a social science context will get search term suggestions from the thesaurus that are semantically related to the initial query such as ``labor market'' or ``education measure''. Another possible suggestion might be ``adolescent'', which is a controlled term for ``youth''. The suggestions generated by this approach go beyond simple term completion and can support the search experience. %\cite{pennell_implementing_2010}
Since ALLBUS and SOEP do not include any annotated entries, we had to use a different corpus to train our term recommender. ALLBUS, SOEP and SOLIS share the same scientific domain and the same domain-specific language which makes this cross-corpora term recommendation plausible.

The term suggestion system generates a ranked list of search term recommendations from which we used the top 20 terms for query expansion. This amount of terms was chosen because the average number of synonyms/related terms of the thesaurus-based query expansion was 19.49.

\section{Results}
\label{sec:results}
We ran a pretest involving simple keyword-based queries generated directly from the log file entries and a hand-crafted query formulation from a domain expert ($Q_{expert}$). Although the results are only slightly better, the domain expert's query formulation is chosen as the baseline for our further experiments. We compare four different automatic QE techniques using two thesaurus-based ($QE_{ot}$ and $QE_{tss}$) and two co-occurrence-based expansions ($QE_{jac}$ and $QE_{cos}$) to this baseline. These different systems are compared to the baseline using R@5, R@10, nDCG@5, and nDCG@10 (see table~\ref{resultTable2}).

\begin{table}[t]
\centering
\caption{Results of the retrieval test on the survey questions comparing four different query expansion (prefix $QE$) techniques to the best manual query formulation technique from a pretest ($Q_{expert}$). Best results are marked in bold font. Statistical significant results are marked with the following confidence level: (*) $\alpha=0.1$.}
\label{resultTable2}
\begin{tabular}{@{}p{2cm}p{1.5cm}p{1.5cm}p{1.5cm}p{1.5cm}@{}}
\toprule
            & R@5       & R@10      & nDCG@5    &nDCG@10\\
\midrule
%$Q_{log}$	& 0.0956	& 0.1339	& 0.4324 &  0.3988\\
$Q_{expert}$& 0.0975    & 0.1502    & 0.4056    & 0.3918\\
$QE_{ot}$	& \textbf{0.1308}	& 0.1854    & 0.4511    & 0.4269\\
$QE_{tss}$  & 0.1245	& 0.1699	& \textbf{0.4857}(*)	& \textbf{0.4486}\\
$QE_{jac}$  & 0.1265	& \textbf{0.1965}	& 0.3411	& 0.3471\\
$QE_{cos}$  & 0.1271	& 0.1938	& 0.4077	& 0.3998\\
\bottomrule
\end{tabular}
\end{table}

In general the thesaurus-based approaches were able to increase both precision and recall. $QE_{ot}$ produces more precise results, while also increasing recall compared to the baseline (nDCG@10 + 8.96\%, R@10 + 23.44\%). An expansion with TSS improves retrieval results both in precision and recall as well (nDCG@10 + 14.5\%, R@10 + 13.12\%). As the results show, $QE_{ot}$ produces a better recall than $QE_{tss}$, while the latter is more precise.
The co-occurrence-based approaches produce similar results compared to each other regarding recall, which is also better than the baseline as well as both thesauri (e.g. $QE_{cos}$: R@10 + 29.03\%). The results of $QE_{cos}$ are slightly more precise than the baseline, though less precise than the results of the thesauri. $QE_{jac}$ produces the lowest values for nDCG.  

Although the results are not statistically significant, the previously mentioned criteria support the validity of the following results: Regarding the thesaurus-based approaches, both thesauri produce distinctly more precise results as well as a greater amount of relevant documents compared to the baseline. While the nDCG values for both co-occurrence-based approaches are better than the baseline, nDCG values are lower. A positive effect of all QE approaches was a higher amount of relevant retrieved question items. All systems return a higher number of relevant question items, while precision is also increased (except for $QE_{jac}$).
With co-occurrence-based expansion, best results are produced with cosine similarity. Cosine similarity delivers a higher number of relevant survey questions, but with less precision compared to the thesauri methods. The most relevant results are retrieved with logarithmic Jaccard index.

\section{Conclusion and Future Work}
\label{sec:conclusion}
Our evaluation shows that manual keyword-based search ($Q_{expert}$) for survey questions suffers from very low recall rates and that our generated approaches can provide better results without losing precision. This demonstrates that for our use case query expansion is an appropriate approach to support survey question retrieval as it provides better recall and precision. This constitutes an important step to facilitate the reuse of survey questions for questionnaire design in the social sciences.

The two different approaches we evaluated in this paper (thesaurus-based and co-occurrence-based expansion) show different results when we compare them to queries manually formulated by a domain expert. Even though our approaches produce better results than the domain expert, the results are not coherent. Generally speaking, the co-occurrence-based approaches were better in increasing recall (R@10) while the thesaurus-based expansions were better in increasing retrieval quality measured by nDCG. This might be related to the different expansion concepts that underlie the current experiment. The concept-relations in a thesaurus are all intellectually curated and hand-crafted by domain experts while the relations we calculated with our co-occurrence methods are statistical values. The co-occurrences show that there is a statistical relatedness between a term and a concept. On average, the statistical methods were able to retrieve more relevant results – although this higher recall comes at the cost of a lower quality.
In general, the statistical methods were more liberal in suggesting term-concept relations while the thesauri were stricter. This is an observation that is also true for the two different thesauri used. The Open Thesaurus as a common language thesaurus was better suited to improve both recall and retrieval quality while the Thesaurus for the Social Sciences showed to be too strict on higher recall levels (R@10). Taking the operationalization process as part of every questionnaire design into consideration, this is hardly surprising as survey questions in general do not address a discipline-specific community. A broader and less domain-specific thesaurus seems to be the better tool for the specific problem we faced in this study.

Another aspect that can be observed is the fact that automatic query expansion not only achieves a higher recall but precision is also higher than through intellectual expansion. Consequently, even domain experts would have profited from the implementation of an interactive recommendation system which offers term or query suggestions during the query formulation phase. In domain-specific search scenarios, this has proven to increase retrieval performance and user satisfaction \cite{gradmann_novel_2011}. Therefore, the TREC-style evaluation setting of this paper has to be expanded for an interactive information retrieval setting.

In future work, we would like to do further experiments regarding a combination of intellectually and automatically generated query expansions. First experiments show promising results as we could further improve nDCG values, as well as the number of retrieved documents by combining the different approaches. We would like to implement different topic-related query expansion systems and evaluate the effects of using these specialized recommenders on each topic. %If this approach leads to better results, we would like to analyze different classification methods in order to choose the best query expansion approach for a specific topic.

\bibliographystyle{splncs03}
\bibliography{tpdl2015}

\end{document}